\documentclass[12pt]{article}
\usepackage{lipsum}

\usepackage{xr-hyper}
\usepackage{amsthm}
\usepackage{fullpage}
\usepackage{footnote}
\usepackage{tabu}
\usepackage{booktabs}
\usepackage[title]{appendix}
\usepackage{scribe}
\usepackage{caption}
\usepackage{setspace}

\providecommand{\keywords}[1]
{
  \small	
  \textbf{\textit{Keywords---}} #1
}
\usepackage{etoolbox}

\makeatletter
\patchcmd{\@makecaption}
  {\parbox}
  {\advance\@tempdima-\fontdimen2} 
  {}{}
\makeatother 

\usepackage{titling} 

\usepackage{amsmath,bbm,amssymb, amsfonts}
\usepackage{mathtools}
\usepackage{graphicx,psfrag,epsf}
\usepackage{enumitem}
\usepackage{natbib}
\usepackage{url} 
\usepackage{color}
\usepackage{algorithmicx}
\usepackage{booktabs}
\usepackage{siunitx}
\usepackage{array}
\usepackage{rotating}

\usepackage{adjustbox}

\usepackage{algorithm}
\usepackage{algpseudocode}

\usepackage{tikz}

\usepackage{url}

\usepackage{multicol}
\usepackage{xcolor}
\usepackage{algorithm}

\makeatletter
\def\breve{\mathpalette\wide@breve}
\def\wide@breve#1#2{\sbox\z@{$#1#2$}%
	\mathop{\vbox{\m@th\ialign{##\crcr
				\kern0.08em\brevefill#1{0.8\wd\z@}\crcr\noalign{\nointerlineskip}%
				$\hss#1#2\hss$\crcr}}}\limits}
\def\brevefill#1#2{$\m@th\sbox\tw@{$#1($}%
	\hss\resizebox{#2}{\wd\tw@}{\rotatebox[origin=c]{90}{\upshape(}}\hss$}
\makeatletter

\newtheorem{theorem}{Theorem}

\def\indep{{\perp \!\!\! \perp}}

\def\gobblestop#1#2{#1}
\def\killstop{%
	\aftergroup\gobblestop
}

\def\thick#1{\hbox{\rlap{$#1$}\kern0.25pt\rlap{$#1$}\kern0.25pt$#1$}}

\def\smbalpha{\boldsymbol{{\scriptstyle{\alpha}}}}

\def\smbalpha{\widehat{\smbalpha}}

\def\hbar{\bar{ h}}

\def\mybox#1{\vskip1mm \begin{center}
        \hspace{.0\textwidth}\vbox{\hrule\hbox{\vrule\kern6pt
\parbox{.9\textwidth}{\kern6pt#1\vskip6pt}\kern6pt\vrule}\hrule}
        \end{center} \vskip-5mm}
\def\lboxit#1{\vbox{\hrule\hbox{\vrule\kern6pt
      \vbox{\kern6pt#1\vskip6pt}\kern6pt\vrule}\hrule}}

\def\thickboxit#1{\vbox{{\hrule height 1mm}\hbox{{\vrule width 1mm}\kern6pt
          \vbox{\kern6pt#1\kern6pt}\kern6pt{\vrule width 1mm}}
               {\hrule height 1mm}}}

\def\fat#1{\hbox{\rlap{$#1$}\kern0.25pt\rlap{$#1$}\kern0.25pt$#1$}}

\newcolumntype{R}{@{\extracolsep{0.5cm}}r@{\extracolsep{0pt}}}%
\newcolumntype{E}{@{\extracolsep{0.25cm}}c@{\extracolsep{0pt}}}%

\makeatletter
\newcommand{\distas}[1]{\mathbin{\overset{#1}{\kern\z@\sim}}}%
\makeatother

\usepackage{xr-hyper}
\usepackage[colorlinks, linkcolor=blue, anchorcolor=blue, citecolor=blue]{hyperref}

\newtheorem{thm}{Theorem}[section]
\newtheorem{cor}[thm]{Corollary}

\newtheorem{lemma}[theorem]{Lemma}

\newtheorem{example}{Example}

\usepackage[left=1in,right=1in,top=1in,bottom=1in]{geometry}

\makeatletter
\newcommand*{\addFileDependency}[1]{
  \typeout{(#1)}
  \@addtofilelist{#1}
  \IfFileExists{#1}{}{\typeout{No file #1.}}
}
\makeatother

\newcommand{\blind}{1}

\begin{document}

\if1\blind
{
	\title{\bf Estimating causal effects of functional treatments with modified functional treatment policies}
	\author{Ziren Jiang$^{1}$, Erjia Cui$^{1}$, Jared D. Huling$^{1}$\thanks{corresponding author: huling@umn.edu}\\
		$^{1}$Division of Biostatistics and Health Data Science, University of Minnesota \\ [8pt]
	}
	\date{}
	\maketitle
} \fi

\if0\blind
{
	\title{\bf Estimating the causal effect of functional treatment with modified functional treatment policy}
	\date{}
	\maketitle
} \fi


\begin{abstract}
Functional data are increasingly prevalent in biomedical research. While functional data analysis has been established for decades, causal inference with functional treatments remains largely unexplored. Existing methods typically focus on estimating the causal average dose–response functional (ADRF), which requires strong positivity assumptions and offers limited interpretability. In this work, we target a new causal estimand, the modified functional treatment policy (MFTP), which focuses on estimating the average potential outcome when each individual slightly modifies their treatment trajectory from the observed one. A major challenge for this new estimand is the need to define an average over an infinite-dimensional object with no density. By proposing a novel definition of the population average over a functional variable using a functional principal component analysis (FPCA) decomposition, we establish the causal identifiability of the MFTP estimand. We further derive outcome regression, inverse probability weighting, and doubly robust estimators for the MFTP, and provide theoretical guarantees under mild regularity conditions. The proposed estimators are validated through extensive simulation studies. Applying our MFTP framework to the National Health and Nutrition Examination Survey (NHANES) accelerometer data, we estimate the causal effects of reducing disruptive nighttime activity and low-activity duration on all-cause mortality.

\end{abstract}%

\newpage
\setstretch{1.775}
\setlength{\abovedisplayskip}{7pt}%
\setlength{\belowdisplayskip}{7pt}%
\setlength{\abovedisplayshortskip}{5pt}%
\setlength{\belowdisplayshortskip}{5pt}%
\section{Introduction}
\begin{figure*}[ht]
    \centering
    \caption{An illustration of minute-level physical activity (PA) records for 50 randomly sampled individuals in the NHANES study, where the PA trajectories of three individuals are highlighted in color and the remaining trajectories are shown in grey.}
     \includegraphics[width=\textwidth]{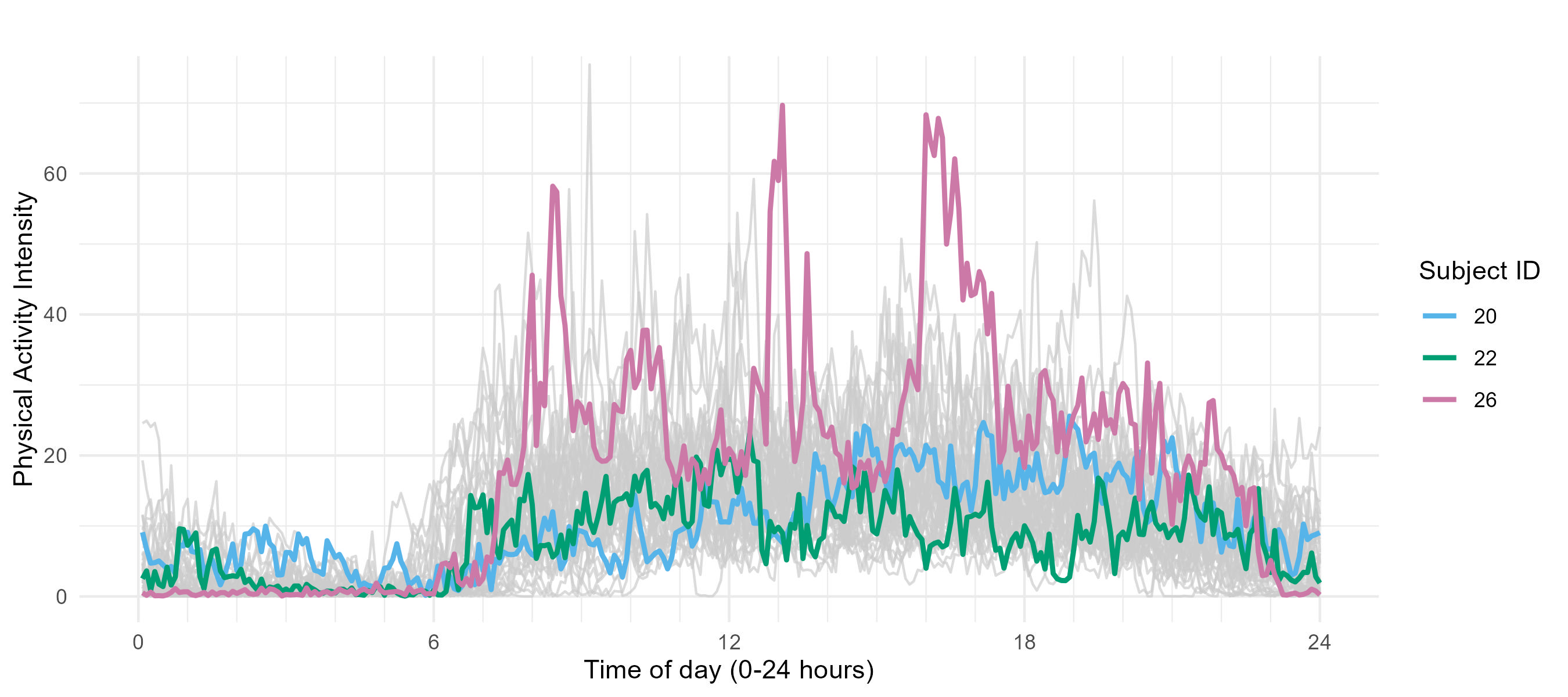}
    \label{Fig:1}
\end{figure*}

The rapid growth of real-time monitoring technologies which track evolving treatments and physiological processes over time has driven the development of statistical methods for functional data. 
In statistics, functional data refer to variables represented as smooth curves or trajectories observed over a continuum such as time, rather than as discrete or scalar measurements. 
One example is the minute-level physical activity trajectory from the National Health and Nutrition Examination Survey (NHANES) \citep{mirel2013national}, with sample trajectories shown in Figure \ref{Fig:1}. Each trajectory implicitly represents an infinite-dimensional curve describing a participant’s physical activity level across an entire day. In many applications, interest goes beyond characterizing functional patterns and focuses on estimating the impact of entire trajectories on clinically relevant outcomes. Specifically, extensive literature in functional data analysis (FDA) has been proposed to study the association between the functional treatment and the outcome \citep{di2009multilevel,cui2021additive,cui2022fast}.

Although these traditional approaches can successfully characterize associations between functional treatments and outcomes, there is growing interest in understanding the causal effects of functional treatments. Yet, estimating causal effects when the treatment itself is a function remains largely unexplored. In this manuscript, we introduce a causal framework for estimating the average potential outcome under a counterfactual world when each individual's treatment trajectory is modified slightly from their observed trajectories.
The choice of the modification policy can be tailored to the scientific question of interest. For example, to investigate the causal effect of reducing nighttime activity on all-cause mortality, our framework allows us to examine expected potential outcomes in a counterfactual scenario in which every individual cuts their nighttime activity intensity to one-half (or one-quarter) of what was observed. If the average potential mortality rate in this counterfactual world is lower, it would indicate a beneficial effect of reducing nighttime activity.

There is a small but growing statistical literature on estimating causal effects of functional treatments. \citet{zhao2025causal} considered functional treatments in the framework of causal mediation analysis and adopted functional regression methods for estimation. \citet{zhang2021covariate}, \citet{tan2025causal}, and \citet{wang2023flexible} all studied the estimation of the causal average dose–response functional (ADRF), defined as the average potential outcome when the entire population receive the same treatment trajectory $A(\cdot)=a(\cdot)$. \citet{zhang2021covariate} estimated the ADRF using a finite-dimensional approximation of the functional treatment obtained via functional principal component analysis (FPCA). \citet{tan2025causal} adopted a functional linear marginal structural model for the ADRF and provided a causal framework along with the estimators. \citet{wang2023flexible}, on the other hand, proposed a weight-modified kernel ridge regression approach to estimate the ADRF, which offers additional modeling flexibility under potential model misspecification.
Despite these contributions, the ADRF has several key drawbacks in the context of functional treatments in practice. First, its identification hinges on a strong positivity requirement: every treatment trajectory $a(\cdot)\in \mathcal{A}$ under consideration must be feasible for all units in the population. This condition is frequently implausible when treatments are infinite-dimensional functions. And because many trajectories are impossible for many individuals, the ADRF itself may be scientifically irrelevant, as it represents the average outcome over the whole population of interest under a specific trajectory. Second, the ADRF itself is an infinite-dimensional functional, which can be challenging to display, summarize, and interpret in applied work. These issues constrain the practical relevance of ADRF-based estimands, which we discuss further in Section~\ref{sec:2} and contrast with our proposed causal estimand.


In addition to the aforementioned literature which views the functional treatment as a single high-dimensional exposure densely measured over a relatively short time window, there is also a rich literature on causal inference in longitudinal studies where treatments are observed at multiple time points, either regularly spaced \citep{ying2023proximal} or irregularly spaced \citep{yang2022semiparametric,roysland2012counterfactual,roysland2025graphical,hu2019causal,rytgaard2022continuous}. As the number of observation times increase to infinity, the longitudinal treatment process can be viewed as an increasingly fine approximation to a functional variable with infinite-dimensionality \citep{ying2024causality}. The key conceptual difference between this longitudinal causal inference framework and the functional causal framework described in the previous paragraph is that the former explicitly accounts for complex temporal confounder–treatment feedback \citep{hernan2010causal}: treatments at earlier times may influence future confounders, which in turn affect subsequent treatments. Within this setting, \citet{ying2024causality} recently established causal identification results that allow for uncountably infinite confounder–treatment feedback. However, they did not develop readily implementable estimators or the corresponding large-sample theory for the proposed estimand.

In this manuscript, we focus on the scenario where the treatment is a ``whole'' functional variable, under the assumption that confounder–treatment feedback does not occur. The rationale is twofold. First, unlike longitudinal studies in which treatments are measured weeks or months apart, many applications involve high-frequency functional data collected over short periods (e.g., minute-level physical activity trajectories measured over a single day in our application). In such cases, temporal confounding feedback within such short intervals may often be negligible. Second, incorporating temporal confounding feedback unnecessarily introduces substantial challenges for estimation and interpretation with infinite dimensionality. Constructing a consistent, let alone root-$n$ consistent, estimator in this setting is generally difficult. To the best of our knowledge, there are no theoretical results regarding this type of estimand.

We introduce a novel causal framework that generalizes the idea of modified treatment policies (MTPs) \citep{haneuse2013estimation, munoz2012population} to settings with functional treatments, which we call modified functional treatment policy (MFTP). MTPs were developed for continuous treatment variables and characterize the causal effect of modifying each individual’s observed treatment according to a prespecified deterministic rule. By anchoring the modification to the observed treatment distribution, MTPs often lead to more plausible positivity conditions and yield causal estimands that can be easier to interpret than those based on the ADRF \citep{sarvet2025natural}. The MTP framework has since been extended to more complex contexts, including longitudinal treatment trajectories \citep{diaz2021nonparametric,hoffman2024studying} and causal mediation analysis \citep{diaz2020causal,gilbert2025identification}, among others. 
As will be demonstrated in Section \ref{sec:2}, our MFTP extension is non-trivial, as it requires defining the population average over a functional random variable whose density function does not exist due to its infinite dimensionality \citep{delaigle2010defining}.  To address this challenge, we first introduce in Section \ref{sec:3} a general and implementable definition of the population average of a functional random variable using its representation through functional principal component analysis (FPCA). We then propose new assumptions on the modification policy that ensure the MFTP estimand is well-defined and causally interpretable. Building on these foundations, we establish causal identification results for the MFTP estimand using weighting and outcome-regression based approaches. Section ~\ref{sec:4} presents corresponding estimators, including a weighting estimator, an outcome regression estimator, and an augmented estimator that combines both models. The asymptotic properties of these estimators are provided in Section ~\ref{sec:5}, where we establish the strong double robustness for the augmented estimator. Section~\ref{sec:6} presents results from simulation studies, and Section~\ref{sec:7} illustrates the proposed methodology by examining the causal effect of daily physical activity on all-cause mortality in the NHANES study.

\section{Causal inference with functional treatment}\label{sec:2}

\subsection{Notation and setup}
Let $A(\cdot)\in\mathcal{A}$ be a random variable representing a functional treatment defined on $\mathcal{T}$, and $\X\in\mathcal{X}\subseteq \mathbb{R}^p$ be a $p$-dimensional vector of pre-treatment covariates. We consider independent and identically distributed data $\{\X_i, A_i(\cdot),Y_i\}$ for subject $i=1,...,n$ where $Y_i$ is the outcome. In practice, we observe a densely measured discrete set of points $\{A_i(t_{i1}),...,A_i(t_{im_i})\}_{i=1}^n$ for the functional treatment variable where $m_i$ is the number of observation points for subject $i$. Denote $Y(a(\cdot))$ as the potential outcome had treatment $a(\cdot)\in \mathcal{A}$ been given.

\subsection{Issues with the causal average dose-response functional}
The existing literature on causal inference with functional treatments has primarily focused on estimation of the causal ADRF, which is defined as
\begin{equation}\label{eq:adrf_defi}
    \mu^{\textnormal{ADRF}}(a(\cdot))\coloneqq 
    \mathbb{E}[Y(a(\cdot))],\;\;\;\;\;\;\;\;\; a(\cdot)\in\mathcal{A},
\end{equation}
where the expectation is taken over the marginal distribution of $Y(a(\cdot))$ for $a(\cdot)\in\mathcal{A}$. The causal ADRF is itself a functional defined in the space $\mathcal{A}$ and considers a counterfactual world where all subjects receive the same exact functional form of the treatment $a(\cdot)\in\mathcal{A}$. The causal ADRF, $\mu^{\textnormal{ADRF}}(a(\cdot))$, is then the average of the potential outcomes under that counterfactual world.

However, the ADRF has several drawbacks, especially when the treatment is a functional variable. First, it requires a stringent positivity/overlap condition to be well defined: the functional treatment $a(\cdot)\in \mathcal{A}$ must be applicable to all subjects in the population, so that there is a positive probability of observing subjects with $A(\cdot)=a(\cdot), \X=\x$ for any $a(\cdot)\in\mathcal{A}$ and $\x\in\mathcal{X}$. However, many treatment patterns are likely to be infeasible to many subjects, making positivity violations likely and the clinical interpretation scientifically irrelevant. For example, in our physical activity case study, a trajectory $a(\cdot)$ representing an intense physical activity (e.g., subject 26 in Figure \ref{Fig:1}) is clearly unlikely to occur for many participants, especially for those with underlying conditions. As the treatment is a functional variable with infinite dimensionality, such an assumption is much less likely to hold.

The second is the interpretability of the ADRF results. With functional treatment, the estimated ADRF becomes a functional that can be difficult to interpret. When some value of treatment $a(\cdot)$ is not applicable to the entire population, the ADRF may lose its scientific relevance altogether. In our physical activity example, consideration of the average potential outcome where all subjects have a physical activity as high as subject 26 is not meaningful for those who are incapable of achieving that level of activity.
Presenting ADRF results can also be challenging. For a scalar continuous treatment, one can plot the ADRF by varying the treatment on the x-axis and the corresponding average potential outcome on the y-axis. With a functional treatment, visualizing the ADRF over $\mathcal{A}$ is nontrivial, making it challenging to communicate with clinicians or domain scientists. 

\subsection{Our contribution: modified functional treatment policies (MFTPs)}
\begin{figure*}[ht]
    \centering
    \caption{An illustration of the potential outcome for the MFTP. The left panel is the observed functional treatment, and the right panel is the modified treatment where we only modified part of the trajectory.}
     \includegraphics[width=0.95\textwidth]{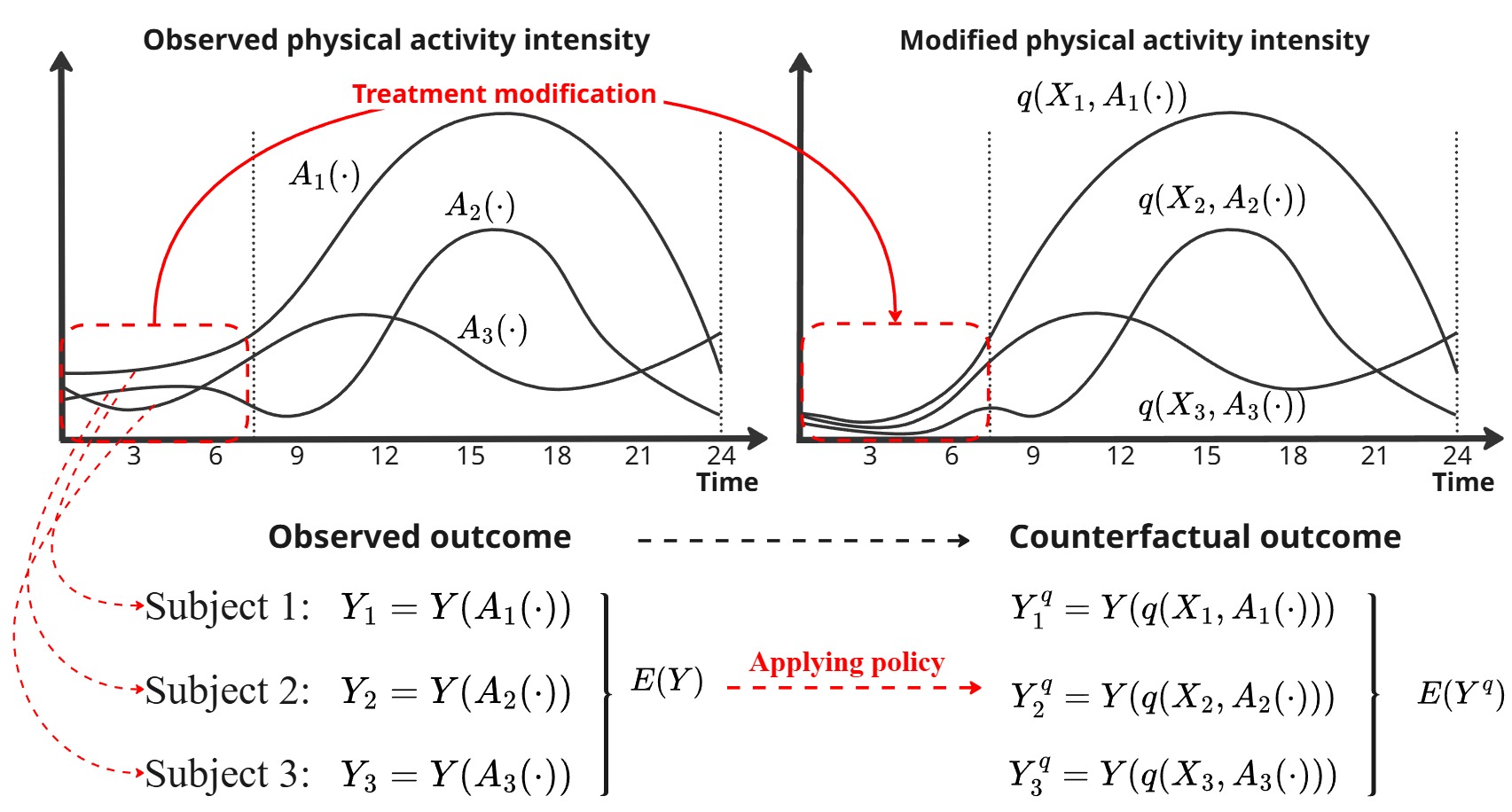}
    \label{Fig:2}
\end{figure*}

In this manuscript, we address the aforementioned concerns of the ADRF by defining a new causal estimand, the modified functional treatment policy (MFTP). Unlike the causal ADRF which assigns the same treatment value to all subjects, MFTPs imagine a counterfactual world where each subject's treatment value is individually adjusted according to $q(\x,a(\cdot))$ where $\x$ is their covariate level and $a(\cdot)$ is their original treatment value (see Figure \ref{Fig:2} for an illustration). The function-valued functional $q(\x,a(\cdot))$ is called the treatment modification policy, which is designed to yield a feasible and interpretable modified treatment trajectory for every subject based on the scientific question of interest.
For example, suppose the treatment is a 24-hour physical activity profile. 
Consider a subject $i$ with underlying conditions characterized by the covariate $\x_i$ whose observed activity 
trajectory is consistently low. 
If we are interested in the causal effect of increasing physical activity, the MFTP framework allows us to define modification policies such as $q(\x_i, a_i(\cdot)) = 1.1 \times a_i(\cdot)$ which slightly increases the amount of activity by $10\%$. Such mild adjustments therefore represent a plausible modification for subject $i$.
The causal estimand in this context is defined as
\begin{equation}\label{eq:mftp_estimand}
    \mu^{q} = \mathbb{E}[Y(q(\X,A(\cdot)))].
\end{equation}
If the treatment variable $A$ is a scalar, \eqref{eq:mftp_estimand} reduces to the definition of a conventional MTP \citep{haneuse2013estimation}. However, it is not trivial to generalize the identification results for conventional MTPs to functional treatments, because the infinite-dimensional functional variable $A(\cdot)$ generally does not have a density function \citep{delaigle2010defining}. Thus, the iterative expectation $\mathbb{E}_{\X, A(\cdot)}\left[\mathbb{E}[Y(q(\X,A(\cdot)))|\X,A(\cdot)]\right]$, which is used to identify the MTP, is not well-defined for functional treatments. In the next section, we overcome this difficulty by proposing a novel definition of the population average over functional variables.

\section{Causal identifiability of MFTPs}\label{sec:3}

\subsection{Population average over functional treatment}\label{sec: 3.1}
A major hurdle for identifying the MFTP estimand is that one cannot simply integrate over a functional object using the conventional definition of expectation. Therefore, we propose to define the population average over functional variables through their functional principal components and show that this coincides with standard definitions of expectations, including that of the observed outcome. 
Denote $\nu(\x,a(\cdot))$ as an arbitrary functional with vector $\x$ and function $a(\cdot)$ as inputs and output a scalar. We aim to define the population average of the functional $\nu$ over the distribution of vector variable $\X$ and functional variable $A(\cdot)$ as
\begin{equation}\label{eq: error_def}
    \mu = \mathbb{E}_{\X,A(\cdot)}[\nu(\X,A(\cdot))].
\end{equation}

Consider the functional principal component analysis (FPCA) that decomposes the variability of the random function $A(\cdot)$ into the spline basis for the space of all square-integrable functions on $\mathcal{T}$:
\begin{equation}\label{eq:fpca}
    A(\cdot) = a_0(\cdot) + \sum_{j=1}^{\infty} \theta_{j}^{1/2}A_j\psi_j(\cdot),
\end{equation}
where $a_0(\cdot)$ is the mean function of $A(\cdot)$ defined as the pointwise mean $a_0(t) = \mathbb{E}[A(t)]$ for all $ t\in \mathcal{T}$, $A_1,A_2,...$ are the uncorrelated random variables of the principal component scores with mean $0$ and variance 1, $\infty>\theta_1\geq\theta_2\geq...$ are the eigenvalues, and $\psi_1(\cdot),\psi_2(\cdot),...$ are the corresponding eigenfunctions \citep{delaigle2010defining}. The existence of such decomposition is guaranteed by the Karhunen-Lo\`{e}ve theorem \citep{ramsay2005functional}. Following \citet{delaigle2010defining}, we assume $A_1,A_2,...$ are independent random variables.

In functional data analysis \citep{morris2015functional}, the functional random variable $A(\cdot)$ can usually be approximated by the first $K$ principal components in FPCA such that $A(\cdot)\approx a_0(\cdot) + \sum_{j=1}^{K} \theta_{j}^{1/2}A_j\psi_j(\cdot)$.
We adopt a similar idea of FPCA approximation to define the population average \eqref{eq: error_def}. We first define the population average up to first $K$ principal components as
\begin{equation}\label{eq:mu_K}
   \begin{split}
        \mu_K &= \mathbb{E}_{\X, A_1,...,A_K}[\nu(\X, a_0(\cdot)+\sum_{j=1}^K\theta^{1/2}_jA_j\psi_j(\cdot))].
   \end{split}
\end{equation}
The quantity $\mu_K$ is well-defined as all the random variables under expectation are scalar. 
We then define the population average of $\nu(\x,a(\cdot))$ over $\X$ and $A(\cdot)$ as the limit of $\mu_K$ when the number of principal components increases to infinity: 
\begin{equation}\label{eq:average_observe}
    \mathbb{E}_{\X,A(\cdot)}[\nu(\X,A(\cdot))] = \mu \coloneqq \lim_{K\to\infty} \mu_K.
\end{equation}
In order for \eqref{eq:average_observe} to be well-defined, we need the following conditions:
\begin{itemize}\label{eq:welldefineassumption}
     \item (\textbf{C1}) For the random function $A(\cdot)$, the eigenvalues of the FPCA, $\theta_1,\theta_2,\ldots$ converge to 0 and have a finite sum, i.e. $\sum_{j=1}^{\infty}\theta_j<\infty$.
     \item (\textbf{C2}) There exists a constant $C_1$ such that, for two square-integrable functions $a_1(\cdot), a_2(\cdot)$ on $\mathcal{T}$, the function $\nu$ in \eqref{eq: error_def} satisfies $|\nu(\x,a_1(\cdot))-\nu(\x,a_2(\cdot))|\leq C_1||a_1(\cdot)-a_2(\cdot)||_2$ for all $\x\in\mathcal{X}$ and $\delta>0$.
     
\end{itemize}
Condition \textbf{C1} is a common assumption for functional data and generally holds for all random functions $A(\cdot)$ that have square integrable kernel \citep{hall2006properties}. It states that the functional principal component analysis can arbitrarily approach the random function $A(\cdot)$ by increasing the number of principal components. Condition \textbf{C2} is a Lipschitz condition for the functional $\nu$ under average, which states that if two functions $a_1(\cdot)$ and $a_2(\cdot)$ are close, then the value of the functionals $\nu(\x,a_1(\cdot))$ and $\nu(\x,a_2(\cdot))$ should also be close at any level of the vector covariates $\x$. Under Conditions \textbf{C1} and \textbf{C2}, the following lemma (with proofs in the Supplementary material) demonstrates that \eqref{eq:average_observe} is well-defined.

\begin{lemma}\textnormal{(Limit of $\mu_K$ exists)} Under conditions \textnormal{\textbf{C1}} and \textnormal{\textbf{C2}}, the sequence of $\{\mu_K\}_{K=1}^{\infty}$ is a Cauchy sequence. Therefore, the limit of $\mu_K$ exists.
\end{lemma}

Equation \eqref{eq:average_observe} defines the average of a functional $\nu$ over the population defined by $\X$ and $A$ using the limit of the average outcome over a finite number of principal components. We further demonstrate that our definition is a natural and intuitive way of representing the population average outcome. We first define the mean conditional outcome:
\begin{equation}\label{eq:m_func}
    \Tilde{m}(\x,a(\cdot)) \coloneqq \mathbb{E}(Y|\X = \x,A(\cdot) = a(\cdot)),
\end{equation}
and $\epsilon\coloneqq Y-\Tilde{m}(\X,A(\cdot))$, where we assume $Y$ to be outcome variable that depends on scalar $\X$ and functional variable $A(\cdot)$. We now have two ways to represent the average outcome, one is through the outcome variable $Y$, where the expectation $\mathbb{E}[Y]$ is naturally defined; the other is through the mean outcome functional $\Tilde{m}(\x,a(\cdot))$, where the average is defined in \eqref{eq:average_observe} with $\nu = \Tilde{m}$. It is natural to expect that they lead to the same ``population average outcome". In fact, the following lemma shows that definition \eqref{eq:average_observe} is equivalent to the population mean of $Y$, which makes this definition a natural one.

\begin{lemma}\label{lemma: gooddefi}
    Under conditions \textnormal{\textbf{C1}}, \textnormal{\textbf{C2}}, and model \eqref{eq:m_func}, we have
    \begin{equation}
        \mathbb{E}_{Y}(Y) = \mu = \lim_{K\to\infty} \mu_K,
    \end{equation}
    where the left-hand side expectation is over the marginal distribution of the scalar random variable $Y$, and $\mu_K$ is defined in \eqref{eq:mu_K} with $\nu=\Tilde{m}$ defined in \eqref{eq:m_func}. 
\end{lemma}

Since the sample mean is a root-$n$ consistent estimator of the expectation, with Lemma \ref{lemma: gooddefi}, we have the following corollary that justifies the sample average estimator of the population average outcome under definition \eqref{eq:average_observe}.

\begin{cor}\textnormal{(Simple consistent estimator of the population average outcome)} Under the assumptions of Lemma \ref{lemma: gooddefi}, the sample mean of the observed outcome $\frac{1}{n}\sum_{i=1}^n Y_i$ is a root-$n$ consistent estimator of the population average outcome defined 
using \eqref{eq:average_observe}.
\end{cor}

\subsection{Causal identification results}\label{sec: 3.2}
In the rest of the manuscript, we adopt the potential outcome language to define the mean outcome functional:
\begin{equation}\label{eq:redefine_m}
    m(\x,a(\cdot)) \coloneqq \mathbb{E}(Y(A(\cdot))|\X = \x,A(\cdot) = a(\cdot)).
\end{equation} 
Similar to \eqref{eq:m_func}, \eqref{eq:redefine_m} defines the mean potential outcome given treatment $a(\cdot)$.
Denote the modified functional treatment variable as $A^q(\cdot) = q(\X, A(\cdot))$. 
Recall that the causal estimand of MFTP is defined as $\mu^q\coloneqq \mathbb{E}\big[Y(A^q(\cdot))\big]$.
To identify $\mu^q$, similar to \eqref{eq:redefine_m}, we first define the conditional mean potential outcome functional under the modification rule $q$ as
\begin{equation*}
    m^q(\x,a(\cdot)) = \mathbb{E}\left[Y(A^q(\cdot))|\X = \x,A(\cdot)=a(\cdot)\right].
\end{equation*}
We further define $\mu_K^q = \mathbb{E}_{\X,A_1,...,A_K}\left[m^q\left(\X, a_0(\cdot)+\sum_{j=1}^K\theta_j^{1/2}A_j\psi_j(\cdot)\right)\right]$. Then, by replacing $Y$ with $Y(A^q(\cdot))$ in Lemma \ref{lemma: gooddefi} and assuming the conditions \textnormal{\textbf{C1}} and \textnormal{\textbf{C2}} with $\nu = m^q$ hold, we have
\begin{equation}\label{eq: muqdoubleexpect}
    \lim_{K\to\infty} \mu^q_K = \mathbb{E}_{\X,A(\cdot)}\big[m^q(\X,A(\cdot))\big] =  \mathbb{E}\big[Y(A^q(\cdot))\big] = \mu^q,
\end{equation}
where $\mathbb{E}_{\X,A(\cdot)}\big[m^q(\X,A(\cdot))\big]$ is defined as in Section \ref{sec: 3.1} and the first two equalities are due to Lemma \ref{lemma: gooddefi}. The following Theorem \ref{the:iden_outcome} proved that $\lim_{K\to\infty} \mu^q_K$ is well-defined.
With our definition of population average over functional variable in \eqref{eq: error_def} and \eqref{eq:average_observe}, results in \eqref{eq: muqdoubleexpect} coincide with the iterative expectation for scalar treatments, which makes it intuitively appealing.

Given the treatment modification rule $q$, the distribution of $A^q(\cdot)$ is uniquely determined by the distribution of $\X$ and $A(\cdot)$. Since the FPCA basis $\{\psi_1(\cdot),\psi_2(\cdot),...\}$ is still an orthonormal basis for the space of all square-integrable functions on $\mathcal{T}$, we can express $A^q(\cdot)$ 
\begin{equation}\label{eq:Aq_defi}
    A^q(\cdot) = a_0(\cdot) + \sum_{j=1}^{\infty} \theta_{j}^{1/2}A^q_j\psi_j(\cdot),
\end{equation}
where $a_0(\cdot)$ is the same mean function for the random treatment $A(\cdot)$ as defined in \eqref{eq:fpca} and $A^q_j, j=1,2,...$ are the random variables corresponding to the scores of the shifted treatment. A subtle point is that the $\theta_{j}$ and $\psi_j(\cdot)$ are still the eigenvalues and eigenfunctions for the observed random function $A(\cdot)$ rather than $A^q(\cdot)$. Therefore, the random variables of new scores $A^q_j, j=1,2,...$ are not guaranteed to be uncorrelated with each other or have variance 1. We thus impose additional conditions to make the estimand well-defined:

\begin{itemize}\label{eq:}
     \item (\textbf{C3} Bounded second moment for the shifted scores): There exists a positive number $C_2$ such that $\mathbb{E}[(A^q_j)^2]\leq C_2$ for all $j$.
     \item (\textbf{C4} Bounded mean and variance for the density ratio): For all $K=1,2,...$, we have $\mathbb{E}\left[\frac{f_{A_1^q,...,A_K^q|\X}(A_1,...,A_K|\X)}{f_{A_1,...,A_K|\X}(A_1,...,A_K|\X)}\right]\leq C_3$ and $\textnormal{Var}\left[\frac{f_{A_1^q,...,A_K^q|\X}(A_1,...,A_K|\X)}{f_{A_1,...,A_K|\X}(A_1,...,A_K|\X)}\right]\leq C_4$, where $A_1,...A_K$ are the scores corresponding to $A(\cdot)$ in \eqref{eq:fpca} and $A_1^q,...,A_K^q$ are the scores corresponding to $A^q(\cdot)$ in \eqref{eq:Aq_defi}.
     \item (\textbf{C5}: Lipschitz condition for the treatment modification rule $q$) There exists a constant $C_5$ such that, for two square-integrable functions $a_1(\cdot), a_2(\cdot)$ on $\mathcal{T}$, the modification policy function $q$ satisfies $||q(\x,a_1)-q(\x,a_2)||_2\leq C_5||a_1(\cdot)-a_2(\cdot)||_2$ for all $\x\in\mathcal{X}$.
\end{itemize}

Condition \textbf{C3} states that the random variables for the scores of the shifted treatment $A^q_j, j=1,2,...$ defined in \eqref{eq:Aq_defi} have bounded second moment. Condition \textbf{C4} states that the density ratio between the MFTP-shifted scores $A^q_1,A^q_2,\ldots$ and the FPCA scores $A_1,A_2,\ldots$ have bounded mean and variance for all values of $K$. Condition \textbf{C5} imposes a Lipschitz-type constraint on the treatment modification rule $q$, ensuring that it does not modify two similar treatment trajectories to substantially different shifted treatments. To make the estimand causally identified, we further impose the following causal assumptions. Denote the modified treatment under policy $q$ as $a^q(\cdot) = q(\x,a(\cdot))$. 

\begin{itemize}\label{eq:}
     \item (\textbf{A1}: Consistency) If $A(\cdot) = a(\cdot)$, then $Y = Y(a(\cdot))$.
     \item (\textbf{A2}: Positivity) If $(\x, a(\cdot))$ belongs to the domain of the functional $m(\x, a(\cdot))$ defined in \eqref{eq:redefine_m}, then $(\x, a^q(\cdot))$ also belongs to the domain.
     \item (\textbf{A3}: Conditional exchangeability) For each $\x,a(\cdot)$, the distribution of $Y(a^q(\cdot))|\X = \x, A = a$ is the same as the distribution of $Y(a^q(\cdot))|\X = \x, A = a^q$.
\end{itemize}

Assumptions \textbf{A1} and \textbf{A2} are common causal assumptions. \textbf{A1} states that the observed outcome is the potential outcome with observed treatment. \textbf{A2} states that for any combination of $(\x, a(\cdot))$ that is realistic for patients (so that it belongs to the domain of $m$), the modified treatment $(\x, a^q(\cdot))$ is also realistic. 
Assumptions \textbf{A3} are generalized from the conditional exchangeability of related populations assumption for MTP. It states that the potential outcome of the modified treatment will not be affected by the original treatment allocation of $a(\cdot)$ or $a^q(\cdot)$. It is less restrictive than the strong ignorability assumption, which states that $Y(a(\cdot))\indep A(\cdot)|\X$ for any $a(\cdot)$ imposed for the ADRF. 
We now present identification results using outcome regression.
\begin{theorem}\label{the:iden_outcome}\textnormal{(Identification through outcome regression)} Under assumptions \textnormal{\textbf{A1}} - \textnormal{\textbf{A3}} and conditions \textnormal{\textbf{C1}}, \textnormal{\textbf{C2}}, and \textnormal{\textbf{C5}}, we have
\begin{equation}\label{eq:outcomeiden}
    \begin{split}
        \mu^q & = \lim_{K\to\infty}\mathbb{E}_{\X,A_1,...,A_K}\left[m\left(\X,q(\X, a_0(\cdot)+\sum_{j=1}^K\theta_j^{1/2}A_j\psi_j(\cdot))\right)\right].
    \end{split}
\end{equation}
\end{theorem}

Theorem \ref{the:iden_outcome} enables us to estimate the causal estimand through estimating the conditional mean outcome $m(\x,a(\cdot))$.
Before presenting the identification results for $\mu^q$ through weighting, we first connect the principal components ($A^q_1,A^q_2,...$) for the MFTP-shifted treatment and the principal components for the original treatment ($A_1,A_2,...$) in terms of the population average through the following lemma. Recall that $\mu_K^q$ is defined as 
\begin{equation*}
    \mu_K^q = \int_{\X,A_1,...,A_K}m^q(\x, a_0(\cdot)+\sum_{j=1}^K\theta_j^{1/2}a_j\psi_j(\cdot))f_{\X,A_1,...,A_K}(\x,a_1,...,a_K) d\x da_1,...,da_K.
\end{equation*}
We now define
\begin{equation*}
    \Tilde{\mu}_K^q \coloneqq \int_{\X,A_1,...,A_K}m(\x, a_0(\cdot)+\sum_{j=1}^K\theta_j^{1/2}a_j^q\psi_j(\cdot))f_{\X,A_1^q,...,A_K^q}(\x,a_1^q,...,a_K^q) d\x da_1^q,...,da_K^q.
\end{equation*}

\begin{lemma}\label{the: Lemma4}
    Under assumptions \textnormal{\textbf{A1}} - \textnormal{\textbf{A3}} and conditions \textnormal{\textbf{C1}} - \textnormal{\textbf{C5}}, we have
    \begin{equation}
        \lim_{K\to\infty} \mu_K^q - \Tilde{\mu}_K^q = 0.
    \end{equation}
    In other words, $\mu_K^q$ and $\Tilde{\mu}_K^q$ are equivalent when $K\to\infty$ and define the same limit $\mu^q$.
\end{lemma}
Lemma \ref{the: Lemma4} indicates that both $\lim_{K\to\infty}\mu^q_K$ and $\lim_{K\to\infty}\Tilde{\mu}^q_K$ define the same population average outcome over the distribution of random functions. The distribution of random functions can thus be fully captured by the distribution of principal component scores when $K$ increases to infinity. 
This allows us to perform dimension reduction when considering the population average over a functional random variable. Based on this connection, we have the following identification result for the causal estimand of MFTP through weighting. 

\begin{theorem}\textnormal{(Identification through weighting)}\label{the:iden_weighting}
    Under assumptions \textnormal{\textbf{A1}} - \textnormal{\textbf{A3}} and conditions \textnormal{\textbf{C1}} - \textnormal{\textbf{C5}}, we have
\begin{equation}\label{eq:weightingident}
    \mu^q = \lim_{K\to\infty} \mathbb{E}\bigg(\frac{f_{A_1^q,...,A_K^q|\X}(A_1,...,A_K|\X)}{f_{A_1,...,A_K|\X}(A_1,...,A_K|\X)} Y\bigg).
\end{equation}
\end{theorem}
Note that $f_{A_1^q,...,A_K^q|\X}$ is the density of MFTP-shifted principal component scores conditional on $\X$. The proof is provided in the Supplementary Material. Theorem \ref{the:iden_weighting} shows that the MFTP effect can be estimated by weighting with the density ratio of the conditional distribution of the principal component scores, effectively reducing the infinite-dimensional functional treatment to a finite-dimensional representation through these scores.

\section{Estimation}\label{sec:4}
We introduce three estimators for the identified MFTP estimand $\mu^q$. Using the identification results, we first construct an outcome-regression (OR) estimator and a weighting (IPW) estimator. We then combine the OR and weighting components to develop an augmented estimator. Theoretical results in the next Section demonstrate its double robustness that attains root-$n$ consistency under weaker conditions than either estimator alone.

\subsection{Outcome regression estimator}
Given an estimator $\hat{m}(\x,a(\cdot))$ of $m(\x,a(\cdot))$, the identification result \eqref{eq:outcomeiden} suggest the following outcome regression estimator:
\begin{equation}\label{eq:OR_estimator}
    \hat{\mu}^q_{\textnormal{OR}} = \frac{1}{n}\sum_{i=1}^n\hat{m}(\x_i,q(\x_i,a_i(\cdot))).
\end{equation}
In practice, the estimator $\hat{m}(\x,a(\cdot))$ can be obtained either through the scalar-on-function regression (SoFR) approaches \citep{reiss2017methods} or other more flexible functional regression models such as deep neural networks (DNN) \citep{wang2024functional}. It is worth noting that although black-box methods for functional data analysis are often criticized for limited interpretability and challenges in uncertainty quantification, they are well-suited in this context since the outcome regression estimator remains interpretable regardless of the interpretability of the adopted outcome model. 

\subsection{Weighting estimator}
For a given number of principal component $K$, denote the true conditional density ratio of the principal components for subject $i=1,...,n$ as
\begin{equation}\label{eq: eki}
    e_{K,i} \coloneqq \frac{f_{A_1^q,...,A_K^q|\X}(A_{1,i},...,A_{K,i}|\X_i)}{f_{A_1,...,A_K|\X}(A_{1,i},...,A_{K,i}|\X_i)}.
\end{equation}
Suppose now we have an estimator $\{\hat{e}_{K,i}\}_{i=1}^n$ of the density ratio function, the identification result \eqref{eq:weightingident} suggests a weighting estimator as
\begin{equation}\label{eq:weighting_estimator}
    \hat{\mu}^q_{\textnormal{IPW}} = \frac{\sum_{i=1}^n\hat{e}_{K,i}Y_i}{\sum_{i=1}^n\hat{e}_{K,i}}.
\end{equation}
In practice, the estimation of \eqref{eq: eki} is equivalent to estimating the balancing weights for MTPs with (multiple) scalar treatment variables. Due to the connection between MTPs and causal average treatment effect on the treated population (ATT) for binary treatments \citep{jiang2025exploring,diaz2021nonparametric}, we can equivalently estimate \eqref{eq: eki} through methods for binary treatment problems with an augmented dataset:
\begin{equation}\label{eq:aug_population}
    \{\widetilde{\X}_i,\widetilde{\A}_i, Z_i\}_{i=1}^{2n}\equiv\{\X_i, \A_i, Z_i = 0\}_{i=1}^{n}\cup\{\widetilde{\X}_{n+i}=\X_{i}, \widetilde{\A}_{n+i}=\A_{i}^q, Z_{n+i} = 1\}_{i=1}^{n},
\end{equation}
where $\A = \{A_1,...,A_K\}$ is the variable of observed first $K$ principal component scores, and $\A^q = \{A^q_1,...,A^q_K\}$ is the random variable of principal component scores for the shifted treatment. $Z$ is the pseudo-treatment indicating membership in the MFTP-shifted population versus the observed population.
We refer to \citet{jiang2025exploring} and \citet{diaz2021nonparametric} for further details in estimating the weights for MTPs, as these methods are directly applicable to this setting.

\subsection{Doubly-robust augmented estimator}
As with many causal estimators, the consistency of the outcome regression estimator \eqref{eq:OR_estimator} and the weighting estimator \eqref{eq:weighting_estimator} depends on the validity of the corresponding estimators $\hat{m}$ and $\hat{e}_K$. In order to provide additional robustness to the model misspecification, we now construct an augmented estimator for $\mu^q$. We demonstrate that it satisfies the ``strong double robustness" property \citep{wager2024causal} in the next section. 

To construct the estimator, we first split the data sample $i=1,...,n$ with index sets $\mathcal{I}_1$ and $\mathcal{I}_2$. Then, for $l\in\{1,2\}$, we estimate $\hat{m}_l$ and $\{\hat{e}_{K,i}^{\mathcal{I}_l}\}_{i\in \mathcal{I}_l}$ within the corresponding data sets $\mathcal{I}_l$. The AIPW estimator can be constructed as
\begin{equation}\label{eq: DResti}
    \begin{split}
        \hat{\mu}^q_{\textnormal{AIPW}} & = \hat{\mu}_{\textnormal{AIPW}}^{q,\mathcal{I}_1} + \hat{\mu}_{\textnormal{AIPW}}^{q,\mathcal{I}_2} \\
        & = \frac{1}{|\mathcal{I}_1|} \sum_{i\in\mathcal{I}_1} \bigg( \hat{m}^{\mathcal{I}_2}(\X_i,A_i^q(\cdot))+\big(Y_i - \hat{m}^{\mathcal{I}_2}(\X_i,A_i(\cdot))\big)\hat{e}_{K,i}^{\mathcal{I}_2}\bigg) \\
        & + \frac{1}{|\mathcal{I}_2|} \sum_{i\in\mathcal{I}_2} \bigg( \hat{m}^{\mathcal{I}_1}(\X_i,A_i^q(\cdot))+\big(Y_i - \hat{m}^{\mathcal{I}_1}(\X_i,A_i(\cdot))\big)\hat{e}_{K,i}^{\mathcal{I}_1}\bigg).
    \end{split}
\end{equation}
We adopt sample splitting to avoid any potential model overfitting and to facilitate the use of more flexible estimation methods while ensuring asymptotic normality. One may consider splitting the sample into more than 2 pieces.


\section{Statistical properties}\label{sec:5}
We now study the statistical properties of the proposed estimators. As we will see in the following theorems, the convergence rate of the estimators not only depends on how well the propensity score of the principal components or the outcome regression model are estimated, but also depends on the properties of the distribution of the random function $A(\cdot)$. 
For a distribution of a random function $A(\cdot)$, let $\theta_1,\theta_2,...$ be the eigenvalues in the functional PCA. Denote $\Delta_K$ as the residual of the eigenvalues: $ \Delta_K\coloneqq \sum_{j=K+1}^{\infty} \theta_j.$
We assume that the number of principal components $K$ increases corresponding to the sample size $n$ in a relationship $K(n)$ (for example, $K = \sqrt{n}$). Then $\Delta_K$ can be viewed as a function of $n$ when $n\to\infty$. In the rest of the manuscript, we replace $K$ with $K(n)$ to emphasize this relationship. We provide several examples on the convergence rate of $\Delta_{K(n)}$ with proof in the Supplementary Material.
\begin{example}\textnormal{\textbf{(Gaussian process with squared-exponential kernel)}} For the case that the random function $A$ follows a Gaussian process with squared-exponential kernel $k(A(t_1),A(t_2))=\exp(-(t_1-t_2)^2/2\sigma_A^2)$, the residual of the eigenvalue follows: 
\begin{equation*}
    \Delta_K=\Theta(B^{K+1})
\end{equation*}
where $\Theta$ refers to the equivalence in terms of order and $0<B<1$ is some constant that does not depend on $K$. Since the rate of $\Delta_{K(n)}$ is exponential in terms of ${K(n)}$, it satisfies $\Delta_{K(n)}=o(n^{-1})$ for any polynomial relationship between ${K(n)}$ and $n$ such as ${K(n)}=n^{1/2}$. 
\end{example}

\begin{example}\textnormal{\textbf{(Gaussian process with Mat\'ern kernel)}} For the case that the random function $A$ follows a Gaussian process with Mat\'ern kernel
\begin{equation*}
    Cov(A(t_1), A(t_2)) = \sigma^{2}{\frac {2^{1-\nu }}{\Gamma (\nu )}}{{\Bigg (}{\sqrt {2\nu }}{\frac {d}{\rho }}{\Bigg )}}^{\nu }K_{\nu }{\Bigg (}{\sqrt {2\nu }}{\frac {d}{\rho }}{\Bigg )}
\end{equation*}
with $d=|t_1-t_2|$ and positive parameters $\nu$ and $\rho$, the residual of the eigenvalue follows: 
\begin{equation*}
    \Delta_K= \Theta(K^{-(2\lfloor\nu\rfloor+1)}).
\end{equation*}
\end{example}

\begin{example}\textnormal{\textbf{(Wiener process)}} For the case that the random function $A$ follows a Wiener process (Brownian motion), the residual of the eigenvalue follows: 
\begin{equation*}
    \Delta_K= \Theta(K^{-1}).
\end{equation*}
In other words, it will never satisfy $\Delta_{K(n)}=o(n^{-1})$ with ${K(n)}<n$. 
\end{example}

In practice, the eigenfunctions are estimated from FPCA with estimation errors \citep{bunea2015sample}. However, the estimated eigenfunctions still form a valid orthonormal basis for the functional space, although they may not minimize the residual variance as the true eigenfunctions do. Therefore, using estimated rather than true eigenfunctions typically do not introduce substantial issues as long as they are reasonably estimated \citep{bunea2015sample}; thus the theoretical results can directly apply to estimated eigenfunctions.
We have the following results for the oracle weighting estimator.
\begin{lemma}\label{lemma:orcleweight}
    For the weighting estimator with oracle weights:
    \begin{equation*}
        \hat{\mu}_e=\frac{1}{n}\sum_{i=1}^{n} e_{{K(n)},i} Y_i,
    \end{equation*}
    under the assumptions in Theorem  \ref{the:iden_weighting} and further assume that: (1) the fourth moments of $A_j$, $e_{K(n)}$, and $e_{K(n)}\epsilon_i$ are bounded for all $j=1,2,...$, ${K(n)}=1,2,...$, and $i=1,...,n$, and (2) the residual of the eigenvalues satisfies  $\Delta_{K(n)} = o(n^{-1})$.
    Then we have $\hat{\mu}_e - \mu^q = O_{p}(n^{-1/2}).$ In other words, $\hat{\mu}_e$ is a root-$n$ consistent estimator.
\end{lemma}

The proof is in the Supplementary Material. With Lemma \ref{lemma:orcleweight}, we have the following results for the convergence rate of the weighting estimator with arbitrary weights $\{\hat{e}_1,...,\hat{e}_n\}$.

\begin{theorem}\label{the: rate_of_weighting}
    For the weighting estimator $\hat{\mu}_{\hat{e}} = \frac{1}{n}\sum_{i=1}^{n} \hat{e}_{{K(n)},i} Y_i $ with weights $\{\hat{e}_1,...,\hat{e}_n\}$, under assumptions in Lemma \ref{lemma:orcleweight}, the estimator is root-$n$ consistent if 
    \begin{equation*}
        \frac{1}{n}\sum_{i=1}^n(\hat{e}_{{K(n)},i}-e_{{K(n)},i})^2 = o_p(n^{-1/2}).
    \end{equation*}
\end{theorem}

In order to demonstrate the statistical properties of the augmented estimator, we first define the oracle AIPW estimator as
\begin{equation}\label{eq: oracleDResti}
    \begin{split}
        \hat{\mu}_{\textnormal{AIPW},n}^*  = \frac{1}{n} \sum_{i=1}^n \bigg( m(\X_i,A_i^q(\cdot))+\big(Y_i - m(\X_i,A_i(\cdot))\big)e_{{K(n)},i}\bigg).
    \end{split}
\end{equation}
The following lemma shows that the oracle AIPW estimator is asymptotically normal.

\begin{lemma}\label{lemma:oracle_aipw} 
    Under assumptions in Theorem \ref{the:iden_weighting} and further assume that: (1) $\textnormal{Var}(\epsilon_i)\leq C_7$ for any $i$, (2) the fourth moments of $ m(\X,A^q(\cdot))+\big(Y - m(\X,A(\cdot))\big)e_{K(n)} -\mu^q$ is finite for any $K(n)$, where 
    $e_{K(n)}=\frac{f_{A_1^q,...,A_K^q|\X}}{f_{A_1,...,A_K|\X}}$. 
    Define $V_{\textnormal{AIPW},K(n)} = \textnormal{Var}[m(\X,A^q(\cdot))+\big(Y - m(\X,A(\cdot))\big)e_{K(n)} -\mu^q]$, then the oracle AIPW estimator satisfies:
    \begin{equation*}
        \frac{\sqrt{n}(\hat{\mu}_{\textnormal{AIPW},n}^* - \mu^q) }{\sqrt{V_{\textnormal{AIPW},K(n)}}} \to N(0, 1).
    \end{equation*}
    
\end{lemma}

The following theorem demonstrates the asymptotic equivalence of the AIPW estimator $\hat{\mu}_{\textnormal{AIPW}}$ with the oracle AIPW estimator under several assumptions. 
\begin{theorem}\label{the: rate_of_DR}
    Under assumptions in Lemma \ref{lemma:oracle_aipw} and further assume (1) $\Delta_{K(n)} = o(n^{-1})$, (2) $\hat{m}^{\mathcal{I}_1}$ and $\hat{m}^{\mathcal{I}_2}$ satisfy the Lipschitz condition as in \textbf{\textnormal{C5}}, (3) the true density ratio $|e_{K(n),i}|\leq C_8$ for all $K$, and (4) the following holds (and also with the roles of $\mathcal{I}_1$ and $\mathcal{I}_2$ swapped),
    \begin{equation}\label{eq: rateassumption}
        \begin{split}
            & n^{2\alpha_{m}} \frac{1}{|\mathcal{I}_1|}\sum_{i\in\mathcal{I}_1} \bigg(\hat{m}^{\mathcal{I}_2}(\X_i,\mathbb{A}_i)-m(\X_i,\mathbb{A}_i)\bigg)^2 \to_p 0 \;\;\;\textnormal{and} \\
            & n^{2\alpha_{e}} \frac{1}{|\mathcal{I}_1|}\sum_{i\in\mathcal{I}_1} \bigg(\hat{e}_{K(n),i}^{\mathcal{I}_2}- e_{K(n),i} \bigg)^2 \to_p 0
        \end{split}
    \end{equation}
    for $\mathbb{A}_i \in \{A_i(\cdot),\;  a_0(\cdot)+\sum_{j=1}^K \theta_j^{1/2}A_{j,i}\psi_j(\cdot), \;a_0(\cdot)+\sum_{j=1}^K \theta_j^{1/2}A^q_{j,i}\psi_j(\cdot)\}$ for all $K$, and 
    for some $\alpha_{m},\alpha_e\geq 0$ and $\alpha_{m}+\alpha_{e}\geq 1/2$.
    
    Then $\sqrt{n}( \hat{\mu}_{\textnormal{AIPW}}- \hat{\mu}_{\textnormal{AIPW}}^*)\to_p 0$ holds, and thus $\sqrt{n}( \hat{\mu}_{\textnormal{AIPW}}- \mu^q) / \sqrt{V_{\textnormal{AIPW},K(n)}}\to N(0, 1).$
    
\end{theorem}

Except for the general assumptions in Theorem \ref{the:iden_weighting}, Theorem \ref{the: rate_of_DR} requires four additional assumptions. The first is that the residual term of the eigenvalues decays faster than $1/n$ in terms of $n$. As illustrated in Lemma \ref{lemma:oracle_aipw}, it is also a required assumption for the estimator with oracle weights to be root-$n$ consistent. The second assumption states that the estimated outcome model is Lipschitz in terms of the functional covariates. This is considered a mild assumption if the outcome model is estimated parametrically, and also reasonable if it is estimated non-parametrically. The third assumption bounds the absolute value of the density ratio. The last assumption is a requirement on the convergence rates of the estimated outcome regression and the density ratio models. However, it does not require either model to be estimated root-$n$ consistently as long as the products of their rates is sufficiently fast.
Theorem~\ref{the: rate_of_DR} demonstrates that our AIPW estimator possesses the \textit{strong double robustness} property; our AIPW estimator achieves root-$n$ consistency even when both the outcome regression and weighting models converge more slowly. As long as their combined convergence rate (i.e., $\alpha_m + \alpha_e$) reaches the root-$n$ threshold, valid inference is still guaranteed. This feature provides substantial practical flexibility, particularly in high-dimensional or functional data settings where estimating either nuisance model at root-$n$ rate is often infeasible.

\section{Simulation experiments}\label{sec:6}
\subsection{Data generating mechanisms and estimation methods}
We use simulation experiments to evaluate the proposed estimators in terms of (1) the empirical consistent rates of the estimators, (2) the coverage rate for the 95\% confidence interval based on non-parametric bootstrap, and (3) the influence of different numbers of principal components $K$ under balancing.
In all the simulation scenarios, we set $T=100$ number of functional observation time points. The functional treatment $\{A_i(t_j)\}_{i=1,...,n; ,j=1,...,T}$ is generated as $A_i(t_j) = a_0(t_j)+\Tilde{A}_i(t_j),$
where $a_0$ is a mean function derived from our data application and $\Tilde{A}$ is a Gaussian process with Gaussian kernel $Cov(\Tilde{A}(t_j), \Tilde{A}(t_{j'})) = e^{\frac{(t_j-t_{j'})^2}{2\sigma_A^2}}$ where $t_j = \frac{j}{T}$ and we vary $\sigma_A\in \{\frac{3}{T},\frac{5}{T}\}$. 
We generate covariates $\mathbf{X}_i$ with $p=15$  for the $i$-th subject with equal proportion from one of the following distributions: 1) normal distribution with mean $\mu_i$ and standard deviation $\sigma_i$, 2) Bernoulli distribution with success probability $p_i$, and 3) Poisson distribution with rate $\lambda_i$. In order to reflect the confounding of the distribution of the covariates and the functional treatment, we set $\mu_i = \textnormal{Average}(\{A_i(t_j)\}_{j=1,...,100})$, $\sigma_i = 1$, $p_i = \frac{\textnormal{Average}(\{A_i(t_j)\}_{j=1,...,10})}{\textnormal{Max}(\textnormal{Average}(\{A_i(t_j)\}_{j=1,...,10}))}$, $\lambda_i = \bigl \lfloor |\textnormal{Average}(\{A_i(t_j)\}_{j=11,...,20})/3| \bigr \rfloor$. 

The outcome variable $Y|\X, A(\cdot)$ follows a normal distribution with mean $\mu_Y$ and variance $1$. The mean outcome $\mu_Y$ follows one of the following relationships: 1) a relatively simple outcome model: $\mu_Y=\eta_A+\eta_X/\sqrt{p},$
where $p$ is the number of scalar predictors
and 2) a more complex outcome model: $\mu_Y= -2(\log{|\eta_A|})^2 + \eta_AX_1+{\eta_X}+\frac{1}{5}X_2X_3^2,$
where $\eta_A = \int A(t)\beta(t)dt$ with $\beta(t)=0.084-(t-0.5)^2$, and $\eta_X = X^T\beta_X$ with $\beta_X$ randomly drawn from $\textnormal{Uniform}(-1,1)$ and fixed across all subjects. We adopt the scalar-on-function regression model $\mathbb{E}[Y] = \int A(t)\hat{\beta}(t)dt + X^T\hat{\beta}_X$ for estimating the outcome regression model
with the state-of-the-art functional data analysis package \texttt{mgcv} in R. Therefore, the simple outcome model represents a scenario without outcome model mis-specification and the more complex outcome model represents a scenario with outcome model mis-specification.

The treatment modification function $q$ is set to be:
\begin{equation*}
    q(\X,A(\cdot))(t) = \tau\times A(t^{1.2}),
\end{equation*}
where $\tau\in \{0.8, 1\}$ modulates the scale of the functional treatment and $A(t^{1.2})$ is a warping function for the modification. As the purpose of the simulation is not comparing different weighting methods which has been studied thoroughly in other causal contexts, we fit the propensity score model with the simplest \texttt{glm} function in R. 

Together with the outcome model, we create four simulation scenarios: (1) no difficulties in either the outcome regression or the weighting (simple outcome model and $\tau=1$), (2) difficulty in weighting (simple outcome model and $\tau=0.8$), (3) difficulty in outcome regression (complex outcome model and $\tau=1$), (4) difficulties in both the outcome regression and weighting (complex outcome model and $\tau=0.8$).

\begin{figure}[ht]
    \centering
    \caption{Simulation results with number of principal components under balancing equals 4. The y-axis is the log of mean squared error and the x-axis is the log of sample size. The scales of y-axis and x-axis are adjusted to be the same.}
     \includegraphics[width=\textwidth]{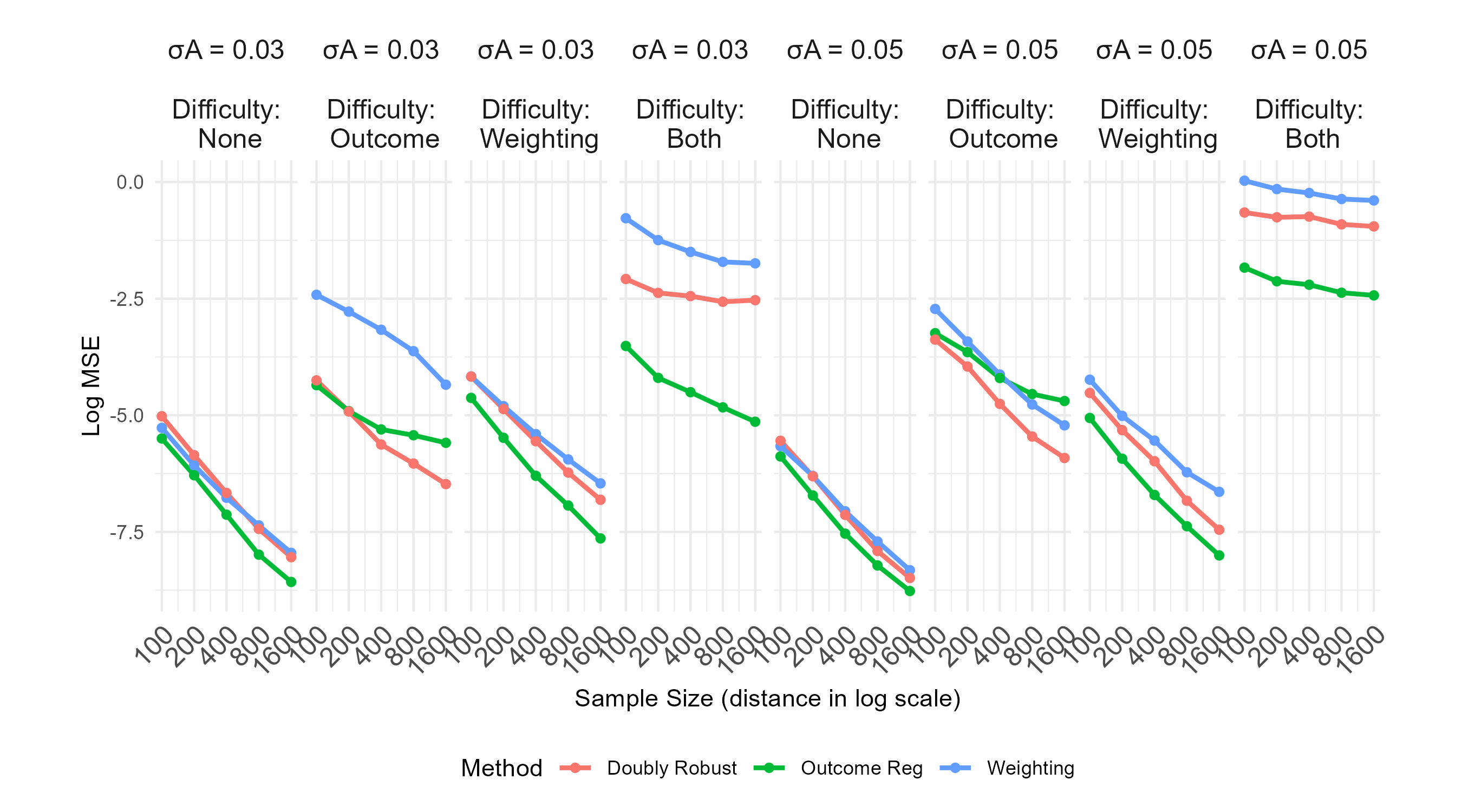}
    \label{Fig:sim1_mse}
\end{figure}

\subsection{Results: convergence rate and coverage}
In the first simulation experiment, we fix the number of principal components under balancing as $K=4$ and vary the sample size $n\in\{100, 200, 400, 800, 1600\}$. We replicate the simulation experiments 1000 times for each scenario and compare the mean squared error (MSE) and the coverage rate with a confidence interval generated from a non-parametric bootstrap. 

The results for the mean squared error are shown in Figure \ref{Fig:sim1_mse}. Figure \ref{Fig:sim1_mse} plots the logarithm of the MSE against the sample size $n$ where the distance is displayed on a log scale. The scales of the x-axis and y-axis are set to be the same. As shown in the figure, the logarithm of the MSE decreases with a slope of $-1$ for the doubly robust estimator in scenarios where at least one of the weighting or outcome regression models is correctly specified. This indicates a root-$n$ performance and the doubly robustness of the estimator. When there are difficulties in both the outcome regression and weighting models, all the estimators have a poor performance and slower than root-$n$ consistency.

\begin{figure}[ht]
    \centering
    \caption{Coverage results for the estimated 95\% confidence interval under various simulation scenarios. The number of principal components under balancing equals 4.}
     \includegraphics[width=\textwidth]{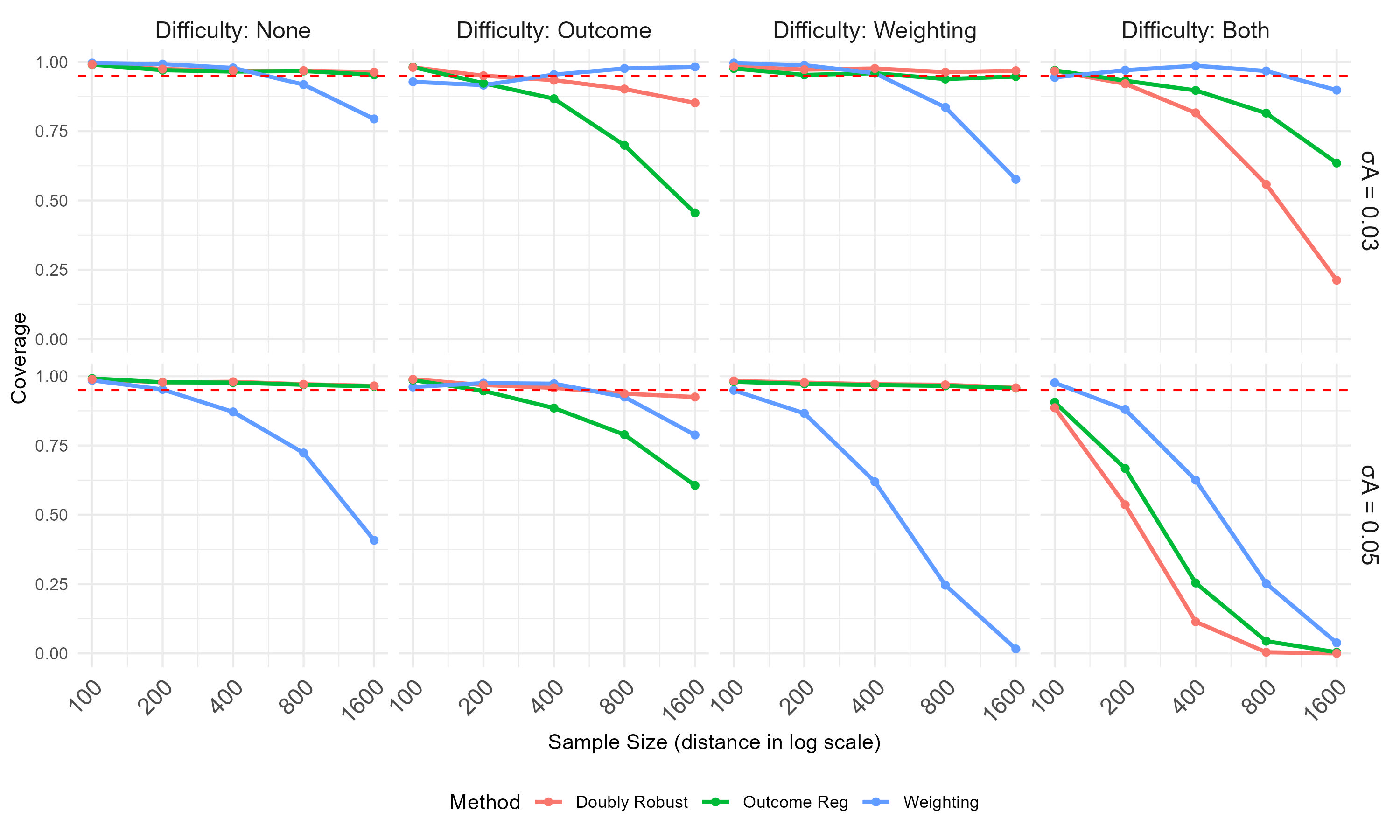}
    \label{Fig:sim1_cover}
\end{figure}

Figure \ref{Fig:sim1_cover} displays the coverage results for the three types of estimators. Similar to the MSE results, the doubly robust estimator has good coverage except for the scenario where there are difficulties in both the outcome regression and weighting models.

\subsection{Results: influence of number of principal components}
\begin{figure}[ht]
    \centering
    \caption{Log of mean squared error for the estimators under different number of principal components.}
     \includegraphics[width=\textwidth]{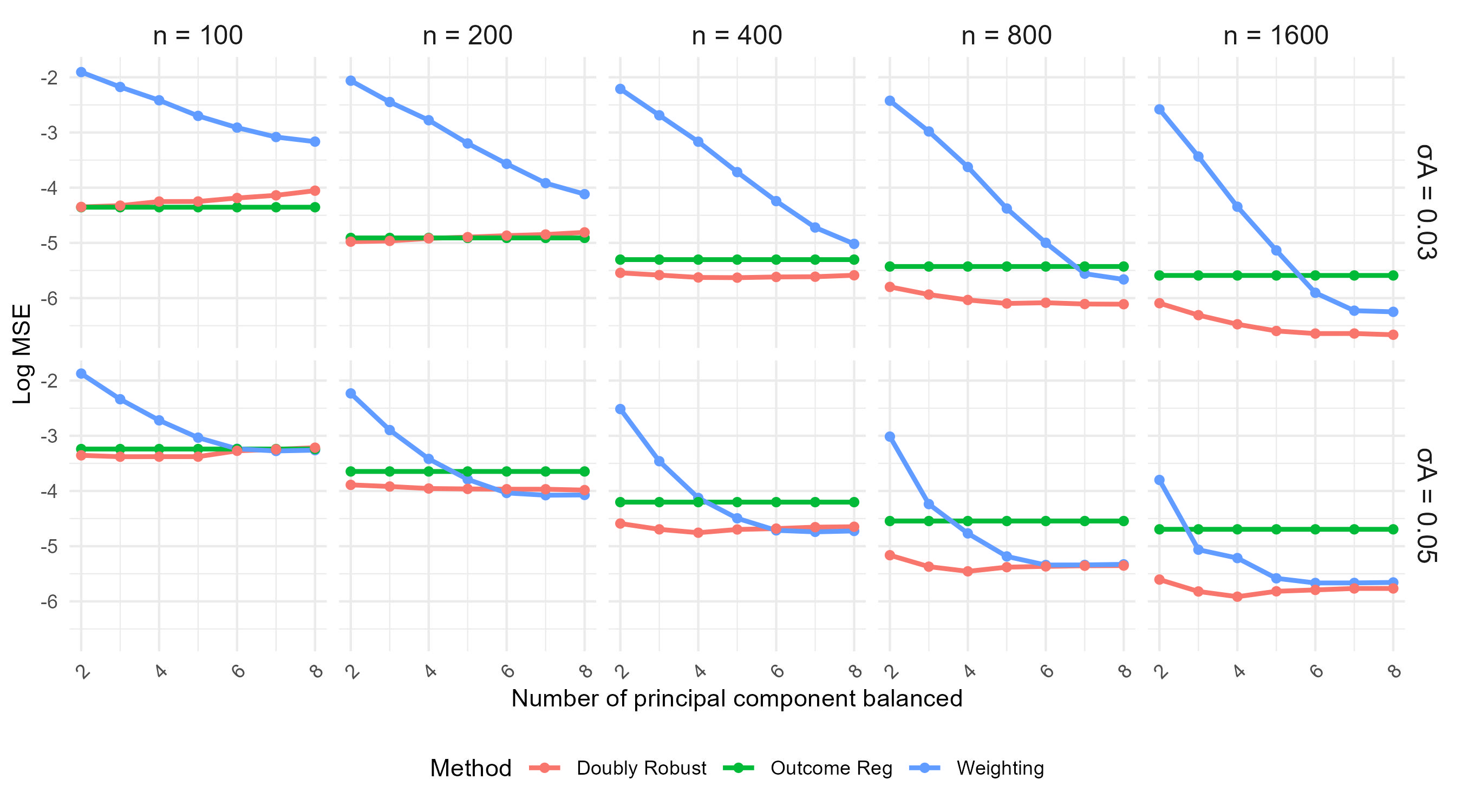}
    \label{Fig:sim2_mse}
\end{figure}

For the second simulation experiment, we fix the data-generating scenario such that there are difficulty in the outcome regression modeling so that the outcome regression estimator does not dominate the result. We vary the number of balanced principal components $K$ from $2$ to $8$. Figure \ref{Fig:sim2_mse} displays the MSE for different $K$ number under the various scenarios. From the figure, the MSE for the weighting estimator generally decreases with an increase of $K$. For the doubly robust estimator, when the sample size is small, increasing $K$ does not help with decreasing the MSE. When the sample size becomes larger, increases in $K$ benefit estimation, though not as substantially as for weighting estimator.


\section{Data application}\label{sec:7}
National Health and Nutrition Examination Survey (NHANES) is a nationally representative study conducted by the U.S. Centers for Disease Control and Prevention (CDC) that evaluates the health and nutritional status of adults and children in the United States through interviews, physical examinations, and laboratory testing \citep{di2017patterns,mirel2013national}. Each wave includes approximately 10,000 participants, with accelerometer-based activity monitoring introduced in recent years as part of a broader effort to capture objective measures of daily behavior. For this case study, we focused on the NHANES 2011–2012 and 2013–2014 cohorts, in which participants were asked to wear a tri-axial wrist-worn accelerometer continuously for up to seven days \citep{leroux2024nhanes}. Wrist-worn accelerometers have become a standard in large-scale epidemiologic studies because they minimize participant burden while allowing continuous, 24-hour monitoring of activity and rest patterns. The raw accelerometer signals were processed by NHANES into the Monitor-Independent Movement Summary (MIMS) units, which provide a device-agnostic metric of physical activity \citep{liu2021assessment}. Following previous studies \citep{cui2021additive,cui2022fast,crainiceanu2024functional}, the accelerometer data are summarized at the minute level, and a log transformation is used to address skewness in the distribution. For each participant, daily activity was averaged across valid wear days, resulting in a total of 1,440 observations for the minute-level average log-MIMS values representing a typical day’s activity profile.

The outcome variable is a binary indicator of five-year all-cause mortality. In addition to functional variable, we include scalar covariates that have been consistently examined in prior studies: age, gender, race/ethnicity, body mass index (BMI), poverty-to-income ratio (PIR), coronary heart disease (CHD), and education level. This covariate set mirrors the adjustment strategy adopted by \cite{crainiceanu2024functional}, who studied the relationship between physical activity and mortality. In this section, we adopt the proposed causal MFTP framework to examine several epidemiological phenomena and provide the corresponding causal insights. 

\subsection{The influence of nighttime activity}
In our first case study, we focus on the influence of disruptive nighttime activity, which is increasingly recognized as an important determinant of some health issues. \citet{zhang2024diurnal} found that irregular sleep was associated with a substantially higher risk of all-cause mortality, independent of total sleep duration and physical activity volume. We aim to explore the causal relationship between excessive nighttime activity and all-cause mortality by applying our MFTP framework to the NHANES dataset.

\begin{figure*}[ht]
    \centering
    \caption{Illustration of our treatment modification policy for the first case study.}
     \includegraphics[width=\textwidth]{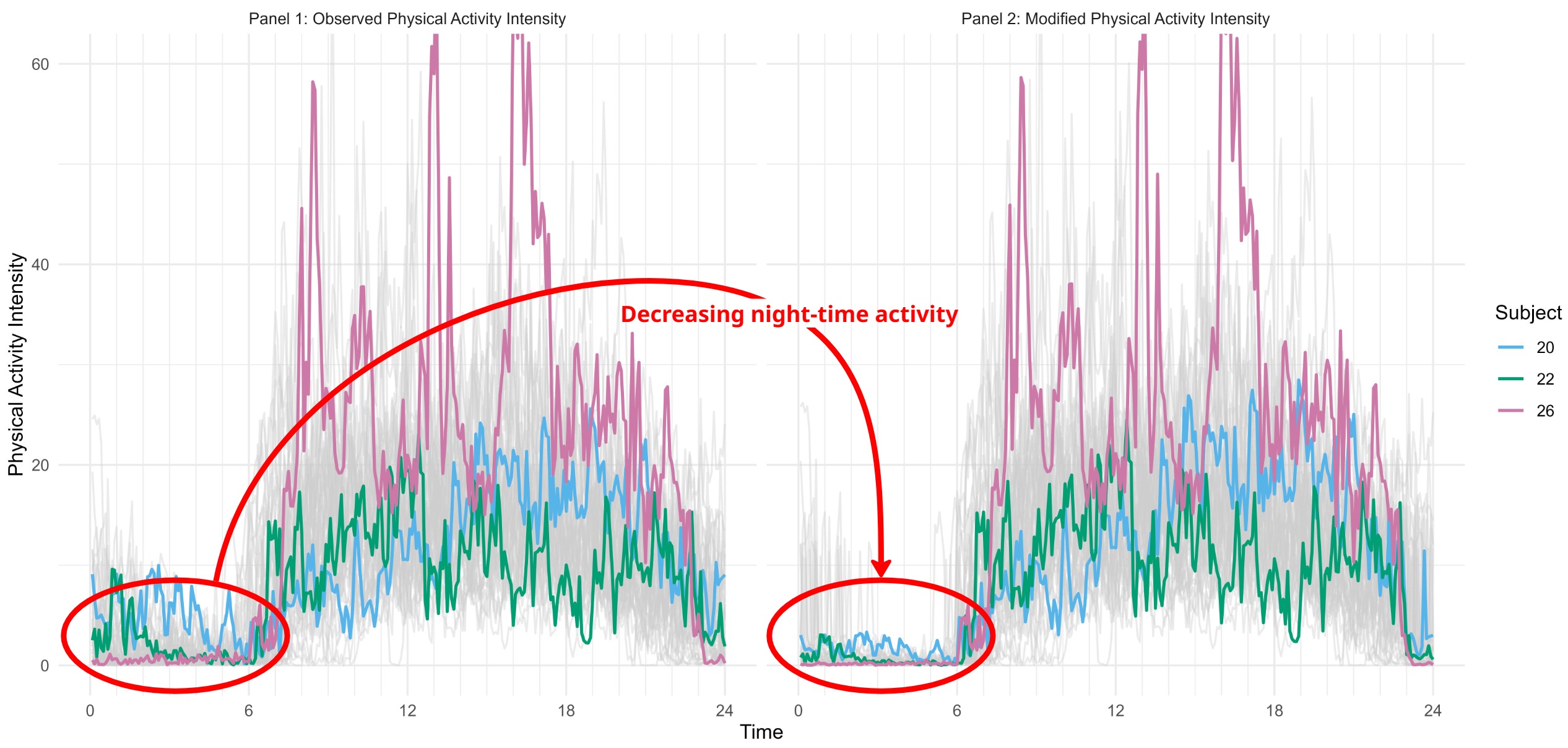}
    \label{fig:casestudy_Intro2}
\end{figure*}

The treatment modification policy for this investigation is defined as
\begin{equation}
    q(\x, a(t)) = 
    \left\{
    \begin{array}{ll}
        c_q \times \tau a(t) & \text{ if } a(t)<10, \\
        c_q \times a(t) & \text{ if } a(t)\geq 10,
    \end{array}
    \right.
\end{equation}
for $t \in [11\ \text{pm}, 6\ \text{am}]$, the parameter $0 < \tau < 1$ controls the magnitude of the modification. To isolate the effect of reducing nighttime activity (rather than reducing the total physical activity), the scaling constant $c_q$ is applied so that the average activity intensity over $\mathcal{T}$ remains unchanged. That is, $c_q$ is defined such that the average activity amount preserves:
\begin{equation}
    \int_{\mathcal{T}} a(t)dt = \int_{\mathcal{T}} q(\x, a(t))dt.
\end{equation}
We restrict the modification to activity levels below 10 units in order to target “disruptive” nighttime activity, while leaving intentional high-intensity nighttime work unaffected. Figure \ref{fig:casestudy_Intro2} illustrates the treatment modification with $\tau = 0.3$ using a subset of the physical activity trajectories. The background gray curves represent 50 randomly sampled trajectories, while three subjects are highlighted in color for illustration. As shown in the figure, the treatment modification rule decreases the disruptive nighttime activity for subjects 20 and 22, while the modification on subject 26 is negligible.

\begin{figure*}[ht]
    \centering
    \caption{Estimated average potential outcome under the different MFTPs. X-axis is the relative ratio of the modified treatment to the observed treatment, where $\tau=0$ represents the largest modification and $\tau=1$ represents no modification.}
     \includegraphics[width=0.9\textwidth]{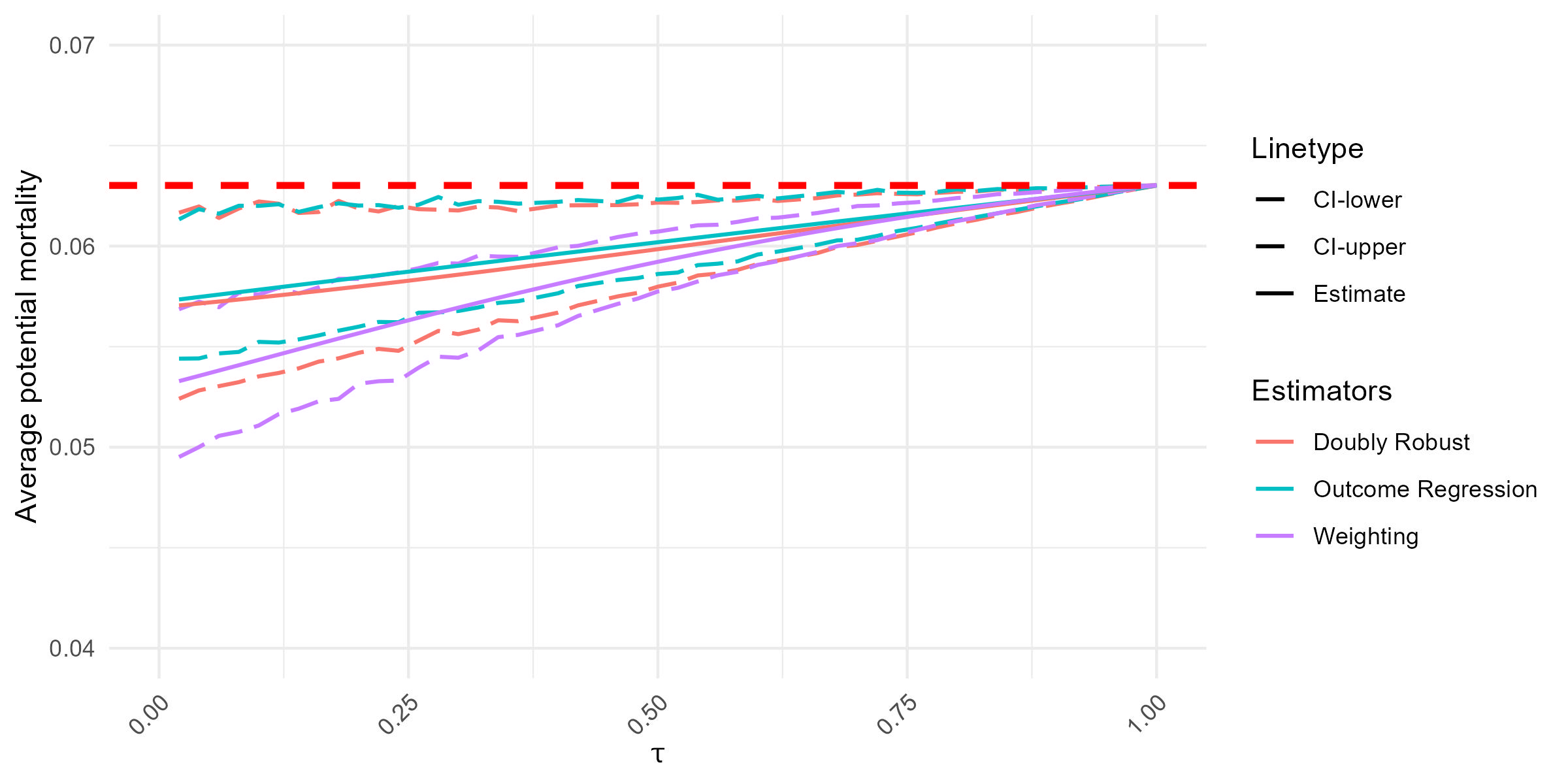}
    \label{fig:casestudy2}
\end{figure*}

We vary $\tau$ between 0 and 1 to represent different levels of modification, and the corresponding results are shown in Figure \ref{fig:casestudy2}. The horizontal axis displays the $\tau$ values ranging from 0 to 1. No modification is applied when $\tau = 1$, resulting in the average mortality equaling the observed mortality rate. As $\tau$ decreases below 1, the average mortality decreases across all estimators, which shows the benefit of reducing the disruptive nighttime activity. Importantly, this result admits a causal interpretation: $\tau$ indexes a well-defined intervention that modifies each individual’s nighttime activity trajectory. The resulting estimand corresponds to the expected mortality under this modified treatment policy, rather than a regression coefficient describing association between observed activity and mortality.

\subsection{The influence of daily life low-activity}
For the second case study, we focus on the excessive low-activity (or sedentary) time which is ubiquitous in modern life. According to \citet{dunstan2012too}, adults are sedentary for around 10 hours per day. A large epidemiologic literature links greater low-activity time with higher risks of cardiovascular disease, type 2 diabetes, and several cancers \citep{di2017patterns,diaz2017patterns,biswas2015sedentary}. However, the majority of the evidence focuses on the ``associative'' relationship rather than ``causative.'' In this subsection, we aim to estimate the causal relationship between the daily life low-activity and all-cause mortality.

\begin{figure*}[ht]
    \centering
    \caption{Illustration of our treatment modification policy for the second case study.}
     \includegraphics[width=\textwidth]{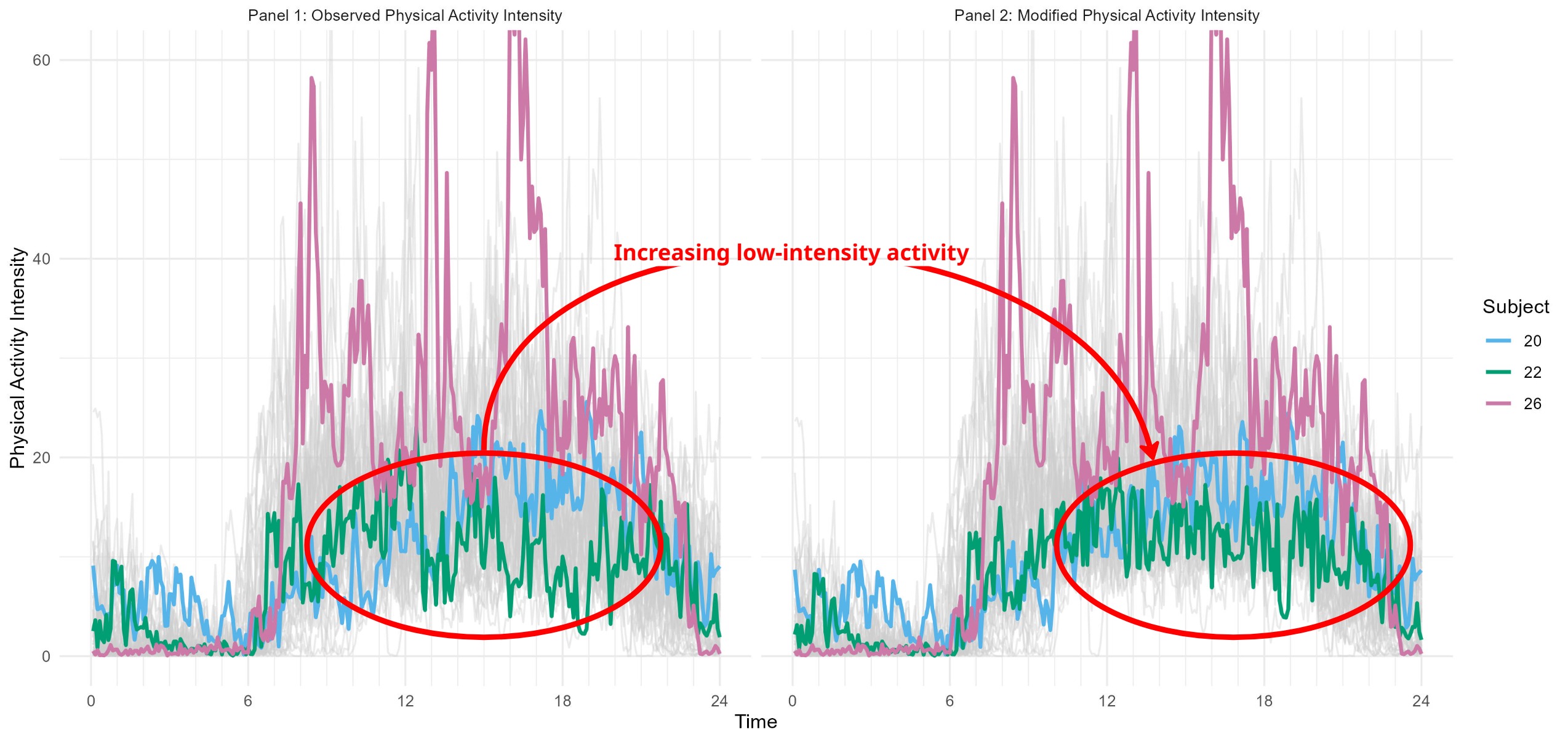}
    \label{fig:casestudy_Intro1}
\end{figure*}

We define low-activity as any minute with an MIMS value below 10 units. The treatment modification policy under consideration is formulated as
\begin{equation}
    q(\x, a(t)) = 
    \left\{
    \begin{array}{ll}
        c_q \times \tau a(t) & \text{ if } a(t)<10, \\
        c_q \times a(t) & \text{ if } a(t)\geq 10,
    \end{array}
    \right.
\end{equation}
for $t\in [10$am to $8$pm$]$ where $\tau > 1$ controls the magnitude of the modification, and $c_q$ is the scaling constant introduced earlier to preserve overall physical activity intensity. The modification increases the activity intensity during the ``sedentary" period, thereby reducing the duration of low-activity bouts. However, to isolate the effect of reducing low-activity (rather than simply increasing total physical activity), we still apply the scaling constant $c_q$ such that the average activity remains unchanged. Figure \ref{fig:casestudy_Intro1} illustrates the treatment modification with $\tau = 2$ using the same subset of the physical activity trajectories as the previous subsection. As shown, the treatment modification rule decreases the ``sedentary" period for subjects 20 and 22. For subject 26, there is no modification because the daytime activity intensity is already generally high without any ``sedentary" period.

\begin{figure*}[ht]
    \centering
    \caption{Estimated average potential outcome under the different MFTPs. X-axis is the relative ratio of the modified treatment to the observed treatment, where $\tau=2$ represents the largest modification and $\tau=1$ represents no modification.}
     \includegraphics[width=0.9\textwidth]{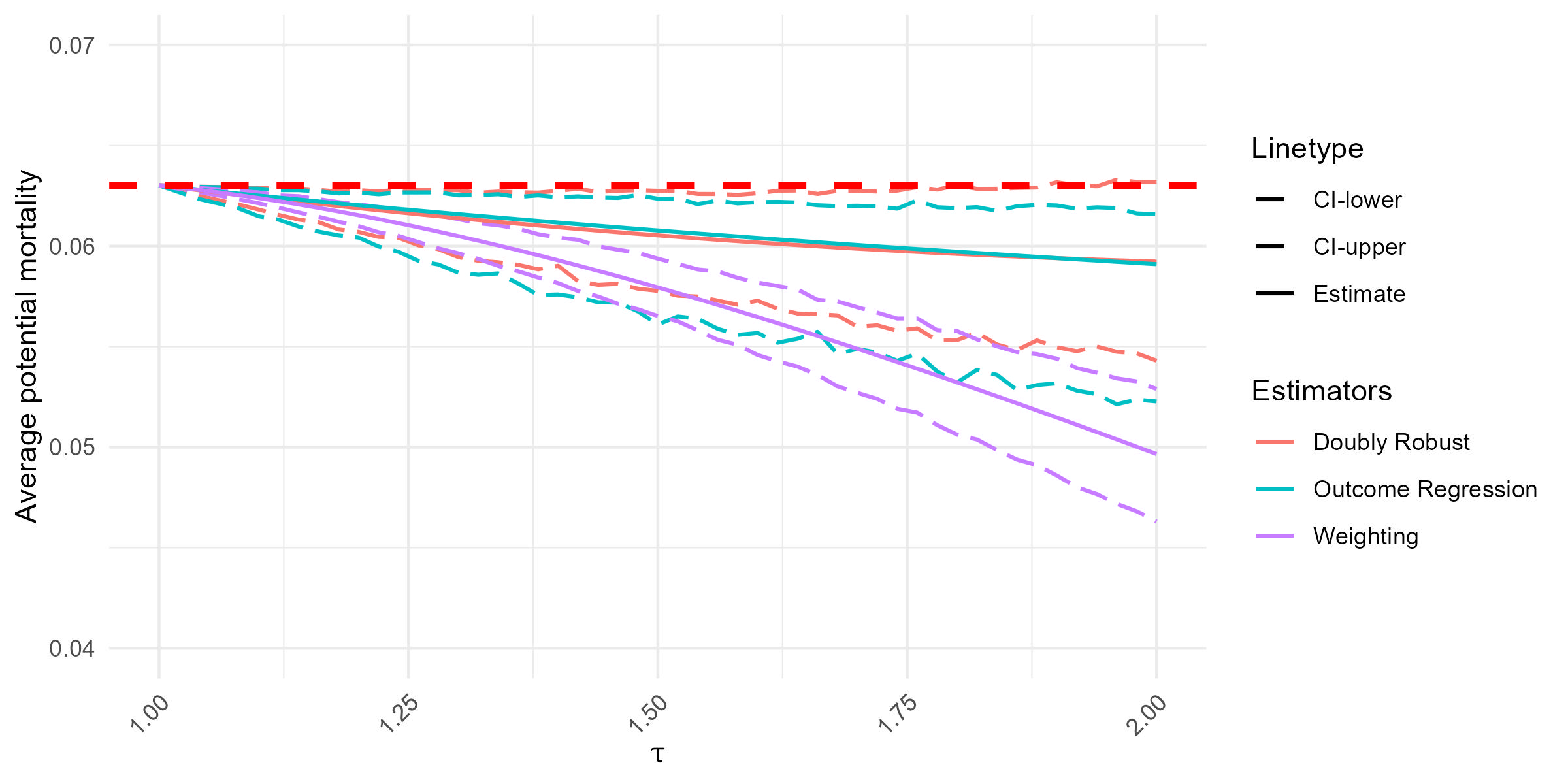}
    \label{fig:casestudy1}
\end{figure*}
We vary $\tau$ between 1 and 2 to represent different levels of modification, and the corresponding results are shown in Figure \ref{fig:casestudy1}. The horizontal axis displays the $\tau$ values ranging from 1 to 2. When $\tau = 1$, no modification is applied to the physical activity trajectories, so the average mortality equals the observed mortality rate. As $\tau$ increases beyond 1, the average mortality decreases across all estimators, demonstrating the benefit of reducing the duration of low-activity periods during the daytime.

\section{Discussion}\label{sec:8}
In this manuscript, we introduce a novel causal framework for functional treatments. The proposed MFTP framework focuses on estimands that imagine a counterfactual world where each subject slightly modifies their treatment, which requires a less stringent positivity assumption compared to estimation of the ADRF. A key methodological innovation lies in defining the population average over functional variables through functional principal component analysis (FPCA), enabling causal identification and estimation using finite-dimensional principal component scores. This approach effectively bridges causal inference and functional data analysis, offering a principled way to handle the infinite-dimensional nature of functional treatments. With the proposed definition, the population average (see Lemma \ref{lemma: gooddefi}) and the conditional expectation (see \eqref{eq: muqdoubleexpect}) for functional variables align with the classical results for scalar variables, making the definition natural and intuitive.

Beyond defining and identifying the MFTP estimand, we further develop weighting, outcome regression, and augmented estimators. Our theoretical results show that the augmented estimator possesses strong double robustness (Theorem \ref{the: rate_of_DR}), allowing it to achieve root-$n$ consistency even if neither the outcome regression model nor the weighting model is estimated at a root-$n$ rate. 
In our data application to the NHANES study, we illustrate through two examples that the proposed MFTP framework enables the investigation of scientifically meaningful questions that would be difficult to address under existing causal frameworks. Compared with the ADRF, our MFTP estimand requires a less stringent positivity assumption and provides results that are more interpretable for the scientific question of interest.

As discussed in the introduction section, our causal MFTP framework considers the scenario where there is no temporal confounding feedback. We demonstrate that it is reasonable for many applications where the functional treatment is measured within a short time window. \citet{ying2024causality,ying2022causal} considered the scenario where there is infinite-dimensional temporal confounding feedback. Future work is needed to develop a corresponding estimator and theoretical justification.





\allowdisplaybreaks


%
%




\def\spacingset#1{\renewcommand{\baselinestretch}%
{#1}\small\normalsize} \spacingset{0.5}


\bibliographystyle{Chicago}
\bibliography{Bibliography}

\end{document}